\documentclass[twoside]{article}
\usepackage{amssymb}
\def\R{{\mathbb R}}

\def\N{{\mathbb N}}

\def\C{{\mathbb C}}

\newtheorem{theorem}{Theorem}

\title{Indefinite metric}
\author{Hanno Gottschalk\\ Institut f\"ur angewandte Mathematik\\
Rheinische Friedrich-Wilhelms-Universit\"at Bonn\\
D-53115 Bonn, Germany\\ gottscha@wiener.iam.uni-bonn.de}

\begin{document}
\maketitle
\pagestyle{myheadings}
\thispagestyle{empty}
\markboth{H. Gottschalk}{Indefinite metric}


\section*{Introduction}
If in a problem of quanitzation state spaces with indefinite inner
product  are used instead of Hilbert spaces, one speaks of
quantization with indefinite metric. The main domain of
application is the quantization of gauge fields, like the
electro-magnetic vector potential $A_\mu(x)$ or Yang-Mills fields
in QCD and the standard model.

The conceptual problem with the indefinite metric is the ocurrence
of senseless negative probablilities in the formalism. Such
negative probabilities however do only arise in expectation values
of fields that are not gauge-invariant and hence do not correspond
do observable quantities. Equivalently, the inner product of
vectors generated by application of such fields to the vacuum
vector with itself can be negative or null. In order to extract
the observable content of an indefinite metric quantum theory, a
subsiduary condition is needed to single out the physical
subspace. Restricted to this subspace, the inner product is
positive semidefinite. This subsiduary condition can bee seen as
the implementation of a gauge, as e.g. the Lorentz gauge
$\partial_\mu A^\mu(x)=0$ in quantum electodynamics (QED). This
procedure is also known under the name Gupta-Bleuler formalism.

The use of indefinite metric in the quantization of gauge theories
like QED can be  avoided all together. This is called quantization
in a physical gauge. The problem with such gauges is that they are
not Lorentz invariant and that the vector potential $A^\mu(x)$ is
not a local field. An example is the Coulomb gauge defined by
$A_0(x)=0$ and $\partial^iA_i(x)=0$ in QED. Furthermore, Dirac
spinor fields $\psi(x)$ in such gauges do not anti-commute when
localized in space-like separated regions. The Dirac fields
therefore also are non-local quantities. Though not in contrast
with special relativity, as Dirac spinors and the vector potential
are not gauge invariant and hence are unobservable, this leads to
severe technical problems in the formulation of interacting
theories. In paricular, the theory of renormalization heavily uses
both - locality and invariance. Therefore, the Gupta-Bleuler
formalism generally is the preferred quantization procedure for a
gauge theory.

That a local and  invariant quantization is not possible using a
(positive metric) Hilbert space has been proven by F. Strocchi in
a series of articles published between 1967 and 1970. If one wants
to preserve locality and/or invariance of the quantized field
theory, it is thus strictly necessary to give up the positivity of
the state space.

A short digression into the early history of the idea might be of
interest. It dates back to 1941, where the use of indefinite
metric in the quantization of relativistic equations was proposed
by Paul Dirac in a price lecture at the
London Royal Society. There, negative probabilities for the
Bosonic vector potential were thought to be connected with the
problem of negative energy solutions of relativistic equations as
a kind of  surrogate of the "Dirac sea" in the quantization of
Fermions. Furthermore, Dirac proposed that negative energy
solutions and negative probabilities would jointly lead to the
cancellation of divergences in QED. The latter idea e.g. was taken
up by W. Heisenberg in his lectures on the theory of elementary
particles held in Munich in 1961, but the generally accepted
solution to the problem of ultra-violet divergences was achived
without recurse to Dirac's original motivation. In 1950 the
consistent quantization of vector potetnial in the  Lorentz gauge
was formulated by S. N. Gupta and K. Bleuler eliminating the use
of negative energy solutions. Indefinite metric since then has
become a building block of the standard theory of quantized gauge
fields.

\section*{No-Go theorems}

The strict necessity of the Gupta-Bleuler procedure for the  local
or covariant quantization of gauge fields has been demonstrated by
F. Strocchi in the form of no-go-theorems for positive metric.
Here we review their content for the case of the electro-magnetic
field. Related statements can be obtained for non-Abelian gauge
theories. The main problem lies in the fact that standard
assumptions on the quantization of relativistic fields are in
conflict with Maxwell equations that should hold as operator
identities in a positive metric theory containing no unobservable
states.
Let 
\begin{equation}
\label{1eqa}
F_{\nu\mu}(x)=\partial_\mu A_\nu(x)-\partial_\nu A_\mu(x)
\end{equation}
 be
the  quantized electro-magnetic field strength tensor. Classically, existence of $A_\mu(x)$ is 
guaranteed from the first set of Maxwell equations $\epsilon^{\alpha\beta\nu\mu}\partial_\beta F_{\nu\mu}(x)=0$. 
Here and in the following indicees are raised and lowered with respect to the Minkowski metric $g_{\alpha\beta}$ and $\epsilon^{\alpha\beta\mu\nu}$ is the
completely antisymmetric tensor on $\R^d$.  Furthermore we apply Einstein's convention on summation over repeated upper and lower indices.	
Standard
assumptions from axiomatic quantum field theory are:
\begin{itemize}
\item[1.] The field strength tensor $F_{\nu\mu}(x)$ is  a operator valued
distribution acting on a (dense core of a) Hilbert space ${\cal
H}$ with scalar product $\langle.,.\rangle$ -- in the indefinite metric case, $\langle.,.\rangle$ only needs to be an inner product;
 \item[2.]
$F_{\mu\nu}(x)$ transforms covariantly, i.e. there
is a strongly continuous unitary (with respect to
$\langle.,.\rangle$) representation $U$ of the orthochronous,
proper Poincar\'e group on ${\cal H}$ such that for translation
$a\in\R^d$ combined with a restricted Lorentz transformation
$\Lambda$ one has
\begin{equation}
\label{2eqa}
U(a,\Lambda)F_{\mu\nu}(x)U(a,\Lambda)^{-1}=(\Lambda^{-1})_{~\mu}^\rho(\Lambda^{-1})_{~\nu}^\kappa
F_{\rho\kappa}(\Lambda x+a);
\end{equation}
 \item[3.] There exists a unique (up to
multiplication with $\C$-numbers) translation invariant vector
$\Omega\in{\cal H}$ (the "vacuum"), i.e. $U(a,1)\Omega=\Omega$
$\forall a\in\R^d$.
\item[4.] The representation of the translations fulfills the spectral condition 
\begin{equation}
\label{3eqa}
\int_{\R^4}\langle\Phi,U(a,1)\Psi\rangle e^{ip\cdot a}\, da=0
\end{equation}
$\forall \Psi,\Phi\in{\cal H}$ if $p$ is not in the closed forward lightcone $\bar V^+=\{p\in\R^4:p\cdot p\geq 0, p^0\geq 0\}$. Here $\cdot$ is the Minkowski inner product.
\end{itemize}
So far the assumptions cocerned only observable quantities. In the following we also demand
\begin{itemize}
\item[5.] The vector potential $A_\mu(x)$ is realized as an operator vaued distribution on ${\cal H}$ and 
transforms covariantly under translations 
\begin{equation}
\label{4eqa}
U(a,1)A_\mu(x)U(a,1)^{-1}=A_\mu(x+a).
\end{equation} 
\end{itemize} 
The assumptions on the nature of the vector poential so far are rather weak. Strocchi's no-go theorems show that one can not add further desirable properties
as Lorentz covariance and/or locality without getting into conflict with the Maxwell equations: 

\begin{theorem} 
\label{1theo}
Suppose that the above assumptions 1.--3. and 5. hold. If Maxwell's equations in the absence of charges
\begin{equation}
\label{5eqa}
\epsilon^{\alpha\beta\nu\mu}\partial_\beta F_{\nu\mu}(x)=0,~~
\partial^\mu F_{\mu\nu}(x)=0
\end{equation}
are valid as a operator identities on ${\cal H}$ and the gauge potential transforms covariantly
\begin{equation}
\label{7eqa}
U(a,\Lambda)A_\mu(x)U(a,\Lambda)^{-1}=(\Lambda^{-1})^\nu_{~\mu}
A_\nu(\Lambda x+a)
\end{equation}
 the two point function of the electro-magnetic field tensor vanishes identically 
\begin{equation}
\label{8eqa}
\langle
\Omega,F_{\nu\mu}(x)F_{\kappa\rho}(y)\Omega\rangle =0~~\forall x,y\in\R^4
\end{equation}
\end{theorem}
To gain a better understanding, where the dificulties in the quantization of the Maxwell equations arise from, here is a rough sketch of the proof:
 Maxwell-equations and
covariance imply that $f_{\mu\nu\rho}(x-y)=\langle\Omega,
A_\mu(x)F_{\nu\rho}(y)\Omega\rangle$ fulfills
$\partial^\alpha\partial_\alpha f_{\mu\nu\rho}(x)=0$ and hence its
Fourier transform has support in the union of the forward and
backward lightcone. The Fourier transform thus can be split into a
positive and a negative frequency part, and
$f_{\mu\nu\rho}=f^+_{\mu\nu\rho}+f^-_{\nu\mu\rho}$ accordingly. By
the general analysis of axiomatic field theory ($\uparrow$), the functions
$f_{\nu\mu\rho}^\pm$ are boundary values of complex analytic
functions on certain tubar domains ${\cal T^\pm}$ transforming
covariantly under a certain representation of the complex Lorentz
group. By a theorem of Araki and Hepp giving a general representation 
of such functions and usig the anti-symmetry of the field tensor, the
following formula can be derived 
\begin{equation}
\label{8eqa}
 f^\pm_{\mu\nu\rho}(z)=(g_{\mu\rho}\partial_\nu-g_{\mu\nu}\partial_\rho)f^\pm(z)
+\epsilon_{\mu\nu\rho\alpha} \partial^\alpha h^\pm(z),~z\in{\cal
T}^\pm
\end{equation}
with $f^\pm,h^\pm$ invariant under complex Lorentz
transformations. Taking boundary values in ${\cal T}^\pm$ one obtains
$f_{\mu\nu\rho}=(g_{\mu\rho}\partial_\nu-g_{\mu\nu}\partial_\rho)f+\epsilon_{\mu\nu\rho\alpha} \partial^\alpha h$
with $f=\bar f^++\bar f^-$ and $h=\bar h^++\bar h^-$ where the bar stands for the distributional boundary value. Maxwell's
equations imply $\partial^\nu f_{\mu\nu\rho}=(\partial^\nu\partial_\nu g_{\mu\rho}-\partial_\mu\partial_\rho) f=0$
and $\epsilon^{~\beta\nu\rho}_\alpha\partial_\beta f_{\mu\nu\rho}=(\partial^\nu\partial_\nu g_{\alpha\mu}-\partial_\alpha\partial_\mu)h=0$. 
The only Lorentz invariant solutions to these equations are constant, which implies the statement of Theorem \ref{1theo}.

The second no-go theorem eliminates the assumption that the vector potential $A_\mu(x)$ is covariant, however a local gauge is assumed. The result is the same
as in Theorem \ref{1theo}:
\begin{theorem}
\label{2theo}
Suppose that the above assumptions 1.--5. and Maxwell's equations hold as operator identities on ${\cal H}$.  If furthermore the gauge is local, i.e. 
\begin{equation}
\label{10eqa}
[A_\mu(x),A_\nu(y)]\linebreak =0 ~~\mbox{if $x-y$ is space-like}
\end{equation}
the two point function of the field strength tensor vanishes again
 as in Theorem \ref{1theo}.
\end{theorem}
Analyzing the interplay of the covariance properties of $F_{\mu\nu}(x)$ with the locaity of $A_\mu(x)$, Strocchi was able to show that the function $f_{\mu\nu\rho}(x-y)$ must have the same covariance 
properties as in Theorem \ref{1theo}, which implies the assertion of Theorem \ref{2theo}. 

The first two no-go theorems deal with the free electromagnetic field that is not coupled to charge carrying fields. This is of course already a real obstruction also for an interacting theory, since by the LSZ formalism
one expects the asymptotic incoming and outgoing fields $A^{\rm in/out}_\mu(x),F^{\rm in/out}_{\mu\nu}(x)$ to be free. In fact, it has been prooven by D. Buchholz that in the positive metric case such asymptotic fields can always be constructed. If
one assumes a local and covariant gauge and positivity, the vanishing of the two-point function would also imply that the field $F_{\mu\nu}(x)=0$ identically by the Reeh-Schlieder theorem.

The next no-go theorem shows that the problems connected to the qantization of the Maxwell equations are not only connected to the free electro-magnetic fields. Let us assume that
the second set of Maxwell equations are given by
\begin{equation}
\label{11eqa}
\partial^\mu F_{\mu\nu}(x)=j_\nu(x)
\end{equation}     
where $j_\nu$ is the leptonic current, i.e. $j_\nu(x)=e:\psi^\dagger(x)\gamma_\nu\psi(x):$ in the case of QED, where $\psi$ is the quantized Dirac field associated to electrons and positrons. Here $:\cdot:$ stands for Wick ordering and $\gamma_\nu$ are the Dirac matrices, $\psi^\dagger=\psi^*\gamma^0$. The conservation on
 the current $\partial^\nu j_\nu(x)=0$ implies that the current charge
\begin{equation}
\label{12eqa}
Q_C=\lim_{R\to\infty}\int_{\R^3} \int_\R\alpha(x^0)\chi(\vec x/R) j_0(x^0,\vec x) ~dx^0d\vec x
\end{equation}
is a constant of motion, where $\alpha$ and $\chi$ are compactly supported infinitely differentiable functions with $\int_\R\alpha(x^0)=1$ and $\chi(\vec x)=1$ for $|\vec x|<1$.  Now, an alternative definition of charge called gauge charge (it generates the global $U(1)$-gauge transformation) is given by 
\begin{equation}
\label{13eqa}
Q_G\Omega=0,~~[Q_C, A_\mu(x)]=0 ~\mbox{and} ~[Q_G,\psi(x)]=-e\psi(x).
\end{equation}
A third formulation of charge, the Maxwell charge $Q_M$ can also be given by replacing $j^0(x)$ in (\ref{12eqa}) by $\partial_\nu F^{\nu0}(x)$. Obviously, if Maxwell equations hold as operator identities, $Q_C=Q_M$. On observable states, all charges $Q_M,Q_C$and $Q_G$ ought to coincide. 
Strocchi's third theorem shows that this cannot be achieved within a local gauge:
\begin{theorem}
\label{3theo}
If the Maxwell equations (\ref{10eqa}) hold and the Dirac field $\psi(x)$ is local w.r.t. the electro-magnetic field tensor $F_{\mu\nu}(x)$, i.e.
\begin{equation}
\label{14eqa}
[F_{\mu\nu}(x),\psi(y)]=0~~\mbox{if $x-y$ is space-like}
\end{equation}
then $[Q_M,\psi(x)]=0$, hence $Q_C=Q_M\not=Q_C$.
\end{theorem}
The proof is a simple consequence of the observation that $j_0(x)=\partial^\nu F_{\nu0}(x)=\partial^iF_{i0}(x)$
 is a three-divergence as $F_{00}(x)=0$ by antisymmetry of $F_{\mu\nu}(x)$. Hence
 \begin{eqnarray}
 \label{15eqa}
 [Q_C,\psi(y)]&=&\lim_{R\to\infty}\int_{\R^4}[j_0(x),\psi(y)]\alpha(x^0)\chi(x/R) ~dx^0d\vec x\nonumber\\
 &=&-\lim_{R\to\infty}\int_{\R^4}[F_{i0}(x),\psi(y)]\alpha(x^0)\partial^i\chi(x/R) ~dx^0d\vec x=0
 \end{eqnarray}
since for $R$ sufficiently large the support of $\alpha(x^0)\partial_i\chi(\vec x/R)$ becomes space like separated from $y$.

It noticable that the proof of none of the above theorems does
rely on the  definiteness of the inner procduct. The main clue of the indefinite metric formalism therefore is rather to give up 
Maxwell equations as operator identities. In the usual positive metric formalism, where all states in ${\cal H}$ are physical states, this would not be legitimate.
But in indefinite metric many states are unobservable -- in particular those with negative "norm" $\langle\Psi,\Psi\rangle<0$. On such states we can give up Maxwell equations. 
This is how indefinite metric comes into the game. 

\section*{Axiomatic framework}

The formalism of axiomatic quantum field theory ($\uparrow$) requires a revision in order to cover the case of gauge fields. The necessary adaptations 
have been elaborated by G. Morchio and F. Strocchi, but also earlier work of E. Scheibe and J. Yngvasson played a r\^ole. 

Let $\phi(x)$ be a $V'$-valued quantum field, where $V$ is a finite dimensional $\C$-vector space with involution $*$. The prime stands for the (topological) dual. For the case of QED, $V$ is eight-dimensional containing four dimensions for the vector potential $A_\mu(x)$ and another
four for the Dirac spinors $\psi(x),\psi^\dagger(x)$. 

Such a quantum field can be reconstructed from its vacuum expectation values (Wightman functions) as follows: Let $S_1=S(\R^{4},V)$ be the space of rapidely decreasing functions $f:\R^{4}\to V$ endowed with  the Schwarz topology. Then the Borcher's algebra 
$\underline{\cal S}$ be the free, unital, involutive tensor algebra over ${\cal S}_1$, i.e. $\underline{S}=\C{\bf 1}\oplus_{n\geq 0}{\cal S}_1^{\otimes n}$ with the multiplication induced by the tensor product and involution $(f_1\otimes\cdots \otimes f_n)^*=f_n^*\otimes\cdots\otimes f_1^*$. $\underline{S}$ 
is endowed with the direct sum topology. One can show that any	linear, normalized, continuous functional $\underline W:\underline{\cal S}\to\C$, $W({\bf 1})=1$, is determined by its restrictions $W_n$ to ${\cal S}_1^{\otimes n}$. By the Schwarz kernel theorem, $W_n\in{\cal S}'(\R^{4n},V^{\otimes n})$. Conversely, any such sequence of Wightman distributions $W_n$ determines a $\underline{W}$.

Given a Hermitean Wightman functional $\underline{W}$
 such that $W(\underline{f}^*)=\overline{\underline{W}(\underline{f})}$, 
$\forall \underline{f}\in\underline{S}$, ${\cal L}_{\underline{W}}=\{\underline{f}\in\underline{\cal S}:\underline{W}(\underline{h}^*\otimes\underline{f})=0~\forall\underline{h}\in\underline{\cal S}\}$ 
forms a left ideal and the inner product $\underline{W}(\underline{f}^*\otimes\underline{h})$ 
induces a non-degenerate inner product $\langle.,.\rangle$ on ${\cal H}_0=\underline{S}/{\cal L}_{\underline{W}}$. Furthermore, the Borchers' algebra $\underline{S}$ acts from the left on ${\cal H}_0$. The quantum field $\phi(x)$ defined as the restriction of this canonical representation to the space ${\cal S}_1\subset\underline{\cal S}$ according to $\phi(f)="\int_{\R^4}\phi^a(x)f_a(x)dx~"$ where the index $a$ runs over a basis of $V$.

If the Wightman functional $\underline{W}$ has further properterties from axiomatic QFT ($\uparrow$) like invariance with resprect to a given representation of the Lorentz group on $V$, translation invariance, locality and the spectral 
property, the quantum field $\phi(x)$ fulfills the related requirements in analogy to the items 1.--5. listed in the previous section for the case of the vector potential $A_\mu(x)$. The Wightman distributions $W_n$ as in the positive metric case are related to the vacuum expectation values of the theory by 
\begin{equation}
W_n^{a_1,\ldots,a_n}(x_1,\ldots,x_n)=\langle\Omega,\phi^{a_1}(x_1)\cdots\phi^{a_n}(x_n)\Omega\rangle.
\end{equation} 
where $\Omega$ is the equivalence class of ${\bf 1}$ in ${\cal H}_0$.

The state space ${\cal H}_0$ produced by the GNS-construction for inner product spaces might be too small to contain all states of physical interest. E.g. in the QED-case it does not contain charged states (cf. Theorem \ref{3theo}).
Depending on the physical problem, one might also be interested to construct coherent or
scattering states and translation invariant states apart from the vacuum. Such states appear in problems related with symmetry breaking and confinement (so-called $\Theta$-vacua) or in some problems of conformal QFT ($\uparrow$) in two dimensions. 
It therefore has become the standard point of view that one needs to take a 
suitable closure of ${\cal H}_0$ such that this closure includes the states of interest (for an alternative point of view see the last paragraph of the following section).   

Typically, larger closures are favourable as they contain more states. One therefore focuses on maximal Hilbert closures of ${\cal H}_0$. A Hilbert topology $\tau$ is induced by an auxilary scalar product $(.,.)$ on ${\cal H}_0$. It is admissible, if it dominates the indefinite 
inner product
$|\langle\Phi,\Psi\rangle|^2\leq C(\Psi,\Psi)(\Phi,\Phi)$ $\forall \Psi,\Phi\in{\cal H}_0$ for some $C>0$. This guarantees that the inner product extends to the Hilbert space closure ${\cal H}$ of ${\cal H}_0$ with respect to $\tau$. Furthermore, there exists a self-adjoint contraction $\eta$ on ${\cal H}$ such that
$\langle\Psi,\eta\Phi\rangle=(\Psi,\eta\Phi)$ $\forall \Phi,\Psi \in{\cal H}$.
A Hilbert topology $\tau$ is maximal, if there is 
no admissible Hilbert topology $\tau'$ that is strictly weaker than ${\cal H}_0$. The classification of maximal admissible Hilbert topologies in terms of the metric operator $\eta$ is given by the following theorem:
\begin{theorem}
\label{3.1theo}
A Hilbert topology $\tau$ on ${\cal H}_0$ generated by a scalar product $(.,.)$ is maximal if and only if the metric operator $\eta$ has a continuous inverse $\eta^{-1}$ on the Hilbert space closure ${\cal H}$ of ${\cal H}_0$. In that case one can replace $(.,)$ by the scalar product
$(\Psi,\Phi)_1=(\Psi,|\eta|\Phi)$ without changing the topology $\tau$. The new metric operator $\eta_1$ then fulfills $\eta_1^2=1_{\cal H}$.
\end{theorem}  
For the proof of the first statement see the original work of G. Morchio and F. Strocchi. One can easily check that $\eta_1=\eta|\eta^{-1}|$ which implies the second assertion of the theorem. A Hilbert space ${(\cal H},(.,.))$ with an indefinite inner product induced by a 
metric operator $\eta$ with $\eta^2=1_{\cal H}$ is called a Krein space. 
 For an extensive study of Krein spaces see the monograph of T. Ya. Azizov and I. S. Iokhvidov.  

One can furthermore show that given a non-maximal admissible Hilbert space topology $\tau$ induced by some $(.,.)$ one obtaines a maximal admissible Hilbert topology as follows: Given the metric operator $\eta$, we define a scalar product $(\Psi,\Phi)_1=(\Psi,(1-P_0)\Phi)$ on ${\cal H}$ with $P_0$ the null space projector of $\eta$. 
Obviously, this scalar product is still admissible and it leads to a new metric operator $\eta_1$ and a new closure ${\cal H}_1$ of ${\cal H}_0$. Furthermore, it is easy to show that the scalar product $(\Psi,\Phi)_2=(\Psi,|\eta_1|\Phi)_1$ still induces a admissible Hilbert topology which is also maximal, as $\eta_2=\eta_1|\eta_1^{-1}|$ clearly fulfills the
Krein relation $\eta_2^2=1_{{\cal H}_2}$.

The question of the existence of a Krein space closure of ${\cal H}_0$ therefore reduces to the question of the existence of an admissible Hilbert topology on ${\cal H}_0$. The following condition on the Wightman functions $W_n$ replaces the positivity axiom in the case of indefinite metric quantum fields: 
\begin{theorem}
\label{3.2theo}
Given a Wightman functional $\underline{W}$, there exists an admissible Hilbert space topology $\tau$ on ${\cal H}_0=\underline{\cal S}/{\cal L}_{\underline{W}}$ if and only if $\exists $ a family of Hilbert seminorms $p_n$on ${\cal S}_n$ such that
$|W_{n+n}(f\otimes h)|\leq p_n(f)p_m(h)$, $\forall n,m\in\N_0$, $f\in{\cal S}_n,h\in{\cal S}_m$.
\end{theorem}

In some cases, covering also examples with non-trivial scattering in arbitrary dimension, the condition of Theorem \ref{3.2theo} canbe checked explicitly, cf. ($\rightarrow$ Indefinite metric: nontrivial models).

It should be mentioned that different choices of the Hilbert seminorms $p_n$ leads to potentially different maximal Hilbert space closures, cf. the articles of G. Hoffmann and Constantinescu/Gheondea. In fact, often the topology is not even Poincar\'e invariant and hence the states that can be approximated with local states depend on a choosen inertial frame. This fact for the case of QED has been interpreted in terms of physical gauges.

Many results from axiomatic field theory ($\uparrow$) with positive metric also hold in the case of QFT with indefinite metric, like the PCT and the Reeh-Schlieder theorem, the irreducibility of the field algebra (for massive theories) and the Bisoniano-Wichmann theorem ($\rightarrow$ algebraic QFT). Other classical results, like the Haag-Ruelle scattering theory and the spin- and statistics theorem definitively do not hold, as has 
been proven by counter examples. This is however far from being a disadvantage, as it e.g. permits the introduction various gauges in the scattering theory of the vector potential $A_\mu(x)$ and fermionic scalar "ghost" fields in the BRST quantization ($\uparrow$) formalism.       
\section*{Gupta-Bleuler gauge procedure}
Here the Gupta-Bleuler gauge procedure is presented in a slightly generalized form following O. Steinmann's monograph. Classically, the equations of motion for the vector potential
$A_\mu(x)$  
\begin{equation}
\label{16eqa}
\partial^\nu\partial_\nu A_\mu(x)+\lambda \partial_\mu\partial^\nu A_\nu(x)=j_\mu(x)
\end{equation}
together with Lorentz gauge condition $B(x)=\partial_\mu A^\mu(x)=0$ imply the Maxwell equations (\ref{11eqa}). Here $\lambda\in\R$ plays the role of a gauge parameter. As seen above, both equations, the so-called pseudo Maxwell equations (\ref{16eqa}) and the Lorentz gauge condition $B(x)=0$, 
can not both hold as operator identities. The idea for the quantization of the theory therefore is to give up the Lorentz gauge condition as an operator identity on the entire state space ${\cal H}$. 

Suppose one has constructed such a theory with an indefinite inner state space ${\cal H}$. Already for the non-interacting theory, any invariant, spectral, local and covariant solution requires indefinite metric, cf. the explicit formula (\ref{17eqa}) below.  To complete the Gupta-Bleuler program, one needs to find a subspace of (equivalence casses of) physical states ${\cal H}'$ of the inner product space ${\cal H}'$ such that the following conditions hold
\begin{enumerate}
\item The vacuum is a physical state, i.e. $\Omega\in {\cal H}'$;
\item Observable fields like $j_\mu(x)$ and $F_{\nu\mu}(x)$ map ${\cal H}'$ to itself;
\item The inner product $\langle.,.\rangle$ restricted to ${\cal H}'$ is positive semidefinite;
\item Observable fields map ${\cal H}''$, the set of null vectors in ${\cal H}'$, to itself;
\item The Maxwell equations hold on ${\cal H}'$ in the sense 
\begin{equation}
\langle\Psi,\partial ^\nu F_{\nu\mu}(x)\Phi\rangle=\langle\Psi,j_\mu(x)\Phi\rangle,~~~\forall \Psi,\Phi\in{\cal H}'.
\end{equation} 
\end{enumerate} 
Then one obtains ${\cal H}^{\rm ph}$ as the completion of the quotient space ${\cal H}'/{\cal H}''$. The physical Hilbert space ${\cal H}^{\rm ph}$ contains the vacuum $\Omega$ (1.), observable fields act on ${\cal H}^{\rm ph}$ (2. and 4.), it is a Hilbert space (3.) and the Maxwell equations hold on it (5.).

To see that such a construction is possible, let us deal with the noninteracting case $j_\nu(x)=0$, i.e. the limit case of vanishing electrical charge $e\to0$, first. By taking the divergence of (\ref{16eqa}) one obtains $(1-\lambda) \partial^\nu\partial_\nu \partial^\mu A_\mu(x)=0$. Excluding Landau gauge  ($\lambda=1$), this implies $(\partial^\nu\partial_\nu)^2A_\mu(x)=0$. 
The most general solution for the two point vacuum expectation values that is in agreement with (\ref{16eqa}) and the requirements of locality, translation invariance, the spectral condition, uniqueness of the vacuum and Lorentz covariance of $A^\mu(x)$ is then
\begin{equation}
\label{17eqa}
\langle \Omega,A_\nu(x)A_\mu(y)\Omega\rangle=(-g_{\mu\nu}+\rho\partial_\mu\partial_\nu) D^+(x-y)+{\lambda\over 1-\lambda}\partial_\mu\partial_\nu E^+(x-y)
\end{equation}
where $D^+$ and $E^+$ are the inverse Fourier transforms of $\theta(p^0)\delta(p^2)$ and $\theta(p^0)\delta'(p^2)$ respectively, where $p^2=p\cdot p$, $\theta$ is the Heavyside function, $\delta$ the Dirac measure on $\R$ of mass one in zero and $\delta'$ its derivative.
$\rho$ and $\lambda$ are gauge parameters, e.g. the Feynman gauge corresponds $\lambda=\rho=0$. We have also omitted an overall factor coresponding to a field strength normalization (choice of numerical value of $\hbar $ -- here $\hbar=1$).

Using Wick's theorem and the GNS-construction for inner product spaces (cf. the preceeding section), it is possible to realize a representation of the vector potential $A_\nu(x)$ as operator valued distribution on some indefinite metric state space ${\cal H}$ with Fock structure, e.g. a Krein closure of the GNS-space with $\Omega$ the GNS-vacuum and ${\cal D}\subseteq{\cal H}$ the canonical domain of definition. In the case of Feynman gauge,
the metric operator $\eta$ can be obtained by a second quantization of the operator $f_\mu\to \sum_{\nu=1}^4 g_{\mu\nu}f_\nu$ on the one-particle space ${\cal S}_1$.

In particular, the field $B(x)$ acts as a operator valued distribution on ${\cal H}$ and from taking the divergence of (\ref{16eqa}) it follows that $\partial^\nu\partial_\nu B(x)=0$. Thus $B(x)=B^+(x)+B^-(x)$ can be decomposed into a positive ("annihilition") and a negative ("creation") frequency part $B^\pm(x)$.  One obtains:
\begin{theorem}
\label{5theo}
The space ${\cal H}'=\{ \Psi\in {\cal D}:B^+(x)\Psi=0\}$ fulfills all requirements 1.--5. of the Gupta-Bleuler gauge procedure.  
\end{theorem}
Condition 1. is obvious and 2. follows from the fact that the fields $F_{\nu\mu}(x)$ and $B(x)$ commute, which can be checked on the level of two-point functions (\ref{17eqa}). In the same spirit, one can also use (\ref{17eqa}) to check 3. and 4. by explicit calculations on the one particle space and showing that ${\cal H}'$ is the Fock space over the one particle states annihilated by $B^+(x)$.
Finally, by Hermiticity of $A^\mu(x)$, $B^+(x)^*=B^-(x)$ and thus $\langle\Psi,B(x)\Phi\rangle=\langle\Psi,B^+(x)\Phi\rangle+\langle B^+(x)\Psi,\Phi\rangle=0$. As the field $B(x)$ stands for the obstruction to Maxwell equations, this implies condition 5..  

It is noticable that the physical state space ${\cal H}^{\rm ph}$ does not depend on the gauge parameters $\lambda,\rho$ and that it is spanned by repeated application of the field tensor $F_{\mu\nu}(x)$ to the vacuum.

By current conservation, the divergence of (\ref{16eqa}) still yields $\partial^\nu\partial_\nu B(x)=0$ also in the interacting case where $e\not=0$. One can then choose the same gauge condition as in Theorem \ref{5theo} to define ${\cal H}'$. One can then try to prove that this space fulfills all the requirements of the Gupta-Bleuler procedure, e.g. in the sense of perturbation theory. Using more advanced formulations from BRST quantization ($\uparrow$) and Bogoliubov's local $S$-matrix formalism, 
this program has been completed up to a solution of the infra-red problem ($\rightarrow$ perturbative renormalization and BRST). 

A different procedure, motivated by the necessity of coincidence of all charges $Q_C,Q_G$ and $Q_M$ on the physical state space, has been elaborated by O. Steinman. It deviates from the standard procedure in the sense that the physical space ${\cal H}'$ is not included in ${\cal H}$, but ${\cal H}^{\rm ph}$ is directly obtained from the GNS-procedure after taking certain limits of Wightman functions restricted to certain gauge-invariant algebras constructed from the Borchers algebra
and a limiting procedure in a gauge parameter. The Wightman functional on this gauge-invariant algebras are positive (in the sense of perturbation theory), the limiting procedure however implies that the so-obtained physical states are singular (i.e. have diverging inner product) to states in ${\cal H}$, hence the so-defined state spaces corresponding to going to a physical gauge {\em after} solving the problem of a perurbative construction of a indefinite metric solution, are not subspaces of  ${\cal H}$.

\end{document}